\documentclass[12pt]{iopart}

\begin{document}

\title[Session A1]{Report on workshop A1: Exact solutions and their interpretation}

\author{Jos\'e M M Senovilla}

\address{F\'{\i}sica Te\'orica, Universidad del Pa\'{\i}s Vasco, Apartado 644, 48080 Bilbao, Spain}
\ead{josemm.senovilla@ehu.es}
\begin{abstract}
I report on the communications and posters presented on exact solutions and their interpretation at the GRG18 Conference, Sydney.
\end{abstract}


\section{Structure of workshop A1}
Workshop A1 received the largest number of submissions in the 18th International Conference on General Relativity and Gravitation held in Sydney, July 2007. This continues a pattern of similar successes in previous GRG conferences. The organization allocated a total of 4 different 2-hour sessions with oral presentations, plus the posters, in order to accommodate the submissions that had been selected. Two of the oral sessions were held on the first conference day (Monday) and the other two on the fourth (Thursday). The total number of oral presentations was 26.

Each 2-hour oral session was devoted to a different subject. To that end, the speakers were grouped by the affinity of their submission contents. Thus, the first session dealt with {\em classical work on exact solutions}, where traditional concepts such as multipole moments, symmetries, Bianchi models, and the Kerr, Szekeres, or Robinson-Trautman solutions were discussed. The second 2-hour session concerned {\em gravitational collapse, spherical symmetry, and charge}. The third session concentrated on the important topic of {\em black holes}, not only in four but also in higher dimensions, so that {\em black rings} and their avatars were an important part of the session. This is a subject of increasing popularity and interest, as many of the traditional properties of 4-dimensional black holes simply do not hold in higher dimensions, and superstring theory may provide some understanding of the underlying properties of black hole thermodynamics. Finally, there was a {\em miscellanea} in the last session, including topics not covered by,  as well as as talks that could not be fitted in, the previous sessions.

In what follows, I report on each of these sessions separately.

\section{Classical work on exact solutions}
The workshop was opened with a short talk by Roy Kerr on the discovery of his fundamental vacuum rotating solution. This talk had a historical, and even sociological, character, summarizing many of the misunderstandings or rumors that have been circulating among the relativity community for so long. The talk was based on the longer and more complete account in \cite{kerr}. Many myths were dispelled, and the central role played by the Robinson-Trautman line-element was highlighted. 

The generalization of the Robinson-Trautman spacetimes to higher dimensions was in fact discussed by \underline{Podolsk\'y}, Ortaggio and \v{Z}ofka. The general spacetime containing a twist-free, shear-free and expanding geodesic null congruence for vacuum, aligned pure radiation or electromagnetic field were presented. In the last case, odd and even dimensions split apart, with a possible additional magnetic field in even dimensions \cite{OPZ}. Surprising and unexpected differences with the 4-dimensional case arise \cite{PO}, mainly: the solutions are of Petrov type D (or 0) exclusively, hence there cannot be gravitational radiation; and there is no analogue to the C-metric. Therefore, the 4-dimensional Robinson-Trautman family is much richer than its higher order generalizations.

Gowdy presented a seemingly new interpretation of the spatially inhomogeneous models with an Abelian group of motions acting on spacelike surfaces of transitivity, which also generate orthogonal surfaces---the so-called generalized Einstein-Rosen waves \cite{Exact}. The key ingredient was the choice of the transitivity surface area element as $G=tr$ in canonical coordinates on the orthogonal surfaces, so that $dG$ can be spacelike, null or timelike in different regions. Actually, these metrics were extensively studied in \cite{BS}, where they were dubbed ``Boost-rotation-symmetric spacetimes''. The spacetimes are claimed to describe gravitational waves travelling in an expanding background given by flat spacetime in Kasner form\footnote{As far as I am aware, there is only one explicit exact solution describing the propagation of gravitational waves in an expanding spatially homogeneous (Bianchi V) background with {\em realistic} matter ---a perfect fluid with $p=\rho/3$---, so that the full non-linear interaction of the expanding matter with the radiation is taken into account. This was discussed at large in \cite{SV}.} \cite{GE}, and to contain ``rogue waves''. Despite some technical difficulties concerning the addition of single pulses at infinity, some well-behaved waves can be constructed. These might have some interest in testing far-field numerical simulations of waves emitted from an isolated source \cite{G}. 

\underline{MacCallum} and Mena discussed the present status of ongoing research concerning the question of when spacetimes with local discrete symmetries turn out to possess a continuous group of motions. The results are as yet unfinished, but there seem to be good chances of improving an old theorem by Schmidt \cite{S}, as well as of proving the existence of an Abelian 2-parameter group of motions acting on the plane of a reflection if the existence of that reflection (on an axis) is assumed.

Work on spatially homogeneous spacetimes, or Bianchi models, was reported by \underline{Van den Bergh}, Karimian and Wylleman. They have proven the remarkable result that geodesic perfect-fluid spacetimes with vanishing divergence of the electric and magnetic parts of the Weyl tensor (relative to the fluid velocity) belong, with some minor purely electric exceptions, to the Class A of Bianchi models (types $I, II, VI_{0}, VII_{0}, VIII, IX$ \cite{Exact}) whenever the shear tensor is degenerate or commutes with the magnetic part of the Weyl tensor \cite{BKVW}. The non-commuting case has not been solved.

B\"ackdahl summarized his recent work on multipole moments. Based on the powerful techniques developed in previous work with Herberthson \cite{BH}, an old-standing conjecture due to Geroch \cite{Ge} was proven under the assumption of analiticity: given any countable set of constants $m_{n}$ such that the power series $\sum_{n=0}^\infty (m_{n}/n!)x^n$ has a positive radius of convergence around $x=0$, there exists a stationary and axially symmetric solution of the vacuum Einstein equations with $m_{n}$ as multipole moments. In addition, the metric can be explicitly reconstructed from the $m_{n}$ following the procedures described in \cite{Ba,BH}.

Bolejko gave a painstaking and very interesting discussion of the Szekeres models \cite{Exact} from a cosmological point of view. These models contain five arbitrary functions of one spatial coordinate (the radius) and no symmetry in general, and therefore constitute a good theoretical lab to test ideas or intuitions about how the actual Universe, which is obviously inhomogeneous at some suitable scales, evolves dynamically in the full non-linear theory. On the one hand, the formation of structures in such expanding spatially inhomogeneous cosmological models was analyzed, with some unforeseen results. By considering double or even triple structures, such as a void close to one or two superclusters of galaxies, the author was able to deduce a growth of density contrast notably faster than in linear or other simplified regimes \cite{Bo}. On the other hand, consequences of the above results in, for instance, the low multipole fluctuations in the spectrum of the cosmic background radiation were considered. The overall conclusion can be stated succinctly as ``local inhomogeneities can affect, notably, the present day cosmological observations.''

\underline{\v{Z}ofka} and Bi\v{c}\'ak reported on their generalization, to the case with a non-vanishing cosmological constant, of the results in \cite{BZ} concerning the (singular) sources of static and cylindrically symmetric vacuum spacetimes. A bound on the mass per unit length of the shell sources can be kept if some extra conditions are enforced.

\section{Gravitational collapse, spherical symmetry, and charge} 
The second session followed after a short break. It was fairly uniform as all speakers treated problems with spherical symmetry. The first presentation was by 
\underline{Lasky} and Lun, who talked on his novel approach to the problem of gravitational collapse with vacuum (Schwarzschild) or null fluid (Vaidya) exterior, including a possible electromagnetic charge \cite{LL}. The main ingredients are two: (i) avoiding the existence of two different unrelated coordinate systems on the interior and exterior metrics, and thereby the use of junction conditions; and (ii) setting up an initial value problem with appropriate matter content up to a finite radius and the appropriate exterior therefrom. This novel approach seems to be powerful and fruitful, and the authors commented on the possibility ---not yet developed--- of including gravitational radiation into the picture by giving up spherical symmetry.

\underline{Krasi\'{n}ski} and Bolejko considered the following question (my wording): is the very last conformal diagram in the Hawking-Ellis book \cite{HE} physically plausible? In other words, the question is whether or not a charged dust can collapse to form a Reissner-Nordstr\"om black hole and nevertheless re-expand into another asymptotically flat region without experiencing any singularity {\em inside} the matter. There is a general negative answer to this question \cite{Ori} if the charge density is strictly smaller than the mass density (in geometric units) everywhere on the dust ball. Apparently, if the former density is larger than the latter, the bounce is possible \cite{KB} but probably not physically realizable. Nevertheless, the physically acceptable case in which the charge density approaches the larger mass density and they eventually become equal at the centre of symmetry had not been contemplated before. This was the possibility discussed in the presentation, and the result turns out to be very surprising: the authors claim that a point-like, instantaneous, curvature singularity at the centre is unavoidable in physically realistic situations \cite{KB}.

Burke, Magesan and \underline{Hobill} gave a genuine exact-solution talk, a fact that is most instructive and encouraging as the speaker introduced himself as a numerical researcher. They have used a simple but powerful method published in \cite{I} to construct solutions of the Einstein-Maxwell equations describing static charged perfect fluids with spherical symmetry. The authors use the particular version of the method in \cite{I} which allows to prescribe the mass density $\rho(r)$ and the charge density $q(r)$ as given functions of the radial (area) coordinate $r$. The linear ODEs can then be solved hoping that the resulting solutions turn out to be physically realistic. The particular Ansatz with $\rho =\rho_{0}-\rho_{1}r^2-\rho_{2}r^4$ and $q=Kr^4$ was chosen, leading to explicit solutions that are fully realistic only when the norm of the static Killing vector can be expressed in terms of trigonometric functions. The equation of state is very complicated but can be plotted numerically showing an approximately linear relation. Charge accumulates at the outer regions of the fluid, and the uncharged limit leads to the realistic Tolman VII solution \cite{T}. In all cases, the extreme Reissner-Nordstr\"om exterior can never describe the exterior of these realistic models. 

\underline{Boonserm} and Visser considered the very old and well-studied problem of static and spherically symmetric perfect fluid solutions. The fact that there only remains a single linear ODE to be solved involving two of the unknown metric functions, and that this ODE can be considered of first or second order depending on the selected unknown function, was used in several (diagonal) coordinate systems to derive ways of ``automatically'' generating solutions from known ones \cite{BV}. The properties of the solutions, specifically concerning regularity at the origin, were analyzed in terms of the seed solution.

\underline{Misthry} and Maharaj considered the matching of a fluid with anisotropic pressures to an exterior Vaidya solution using Gaussian coordinates on the interior. The particular explicit model presented at the talk had already been discussed in \cite{NGG} ---though some linear transformations of coordinates were performed which, of course, do not alter the properties of the solution. A similar problem was discussed by \underline{Sharif} and Ahmad, who also used Gaussian coordinates including now a positive cosmological constant $\Lambda$, but with a static exterior ---the Schwarzschild-Kottler solution \cite{Exact}. The authors studied the properties of the apparent horizons and concluded that $\Lambda>0$ slows down the collapse of matter leading to smaller sizes of possible black holes \cite{SA}.

\section{Black holes and black rings}
The second workshop day started with a session on Black Holes and its higher dimensional relatives. The first talk by \underline{Griffiths} and Podolsk\'{y} gave a comprehensive compendium of accelerating and rotating black holes by considering the different cases contained in the Pleba\'{n}ski-Demia\'{n}ski \cite{Exact} type-D vacuum line-element. By choosing the parameters and coordinates appropriately they were able to describe a generically accelerating and rotating NUT black hole (i.e. with four arbitrary constants $m,a,l,\alpha$ for the mass, rotation, NUT and acceleration parameters, respectively). It turns out that the cases with $l>a$ do not have any curvature singularity, in contrast with those having $a>l$, which contain a ring curvature singularity. This indicates that the natural candidate for an ``accelerating NUT'' solution ($a=0$) may not exist. Actually, one can prove that in this case with $a=0$ the accelaration parameter $\alpha$ can be removed by coordinate transformations leaving the pure NUT solution \cite{GP}.

\underline{Iguchi}, Mishima and Tomizawa presented their work on black holes in Kaluza-Klein bubbles. These objects are known to possess a structure even richer than the higher-dimensional asymptotically flat black holes. By using the B\"acklund transformation or inverse scattering method the authors were able to build vacuum solutions with an Abelian 3-parameter group of motions. The solutions are interpretable as rotating \cite{TIM} or boosted \cite{IMT} pairs of black holes with $S^3$ horizon topology in a 5-dimensional Kaluza-Klein theory. These black holes are kept in balance by a ``Kaluza-Klein bubble'' \cite{W}. The boosted case has a limit with a single boosted black string.
\underline{Mishima} and Iguchi \cite{IMi} also used solitonic generation techniques to superpose several $S^1$-rotating black rings \cite{ER} in a concentric manner. In this case, the solutions are  asymptotically flat. Focusing on the basic superposition of two black rings, called the ``black di-ring'', the parameters can be chosen in such a way that closed timelike lines and conical or curvature singularities are avoided. The remaining freedom is still very large, though, containing four arbitrary constants. Consequently, there is an infinite (2-dimensionally continuous, not discrete) number of black di-rings with the same mass and angular momentum. The so-called fat and thin black rings \cite{ER} are both limits of the black di-ring solution, hence it was argued that one can transform one into the other by means of a continuous path of black di-rings.

The last talk in this session actually dealt with the fundamental problem of black ring stability. Elvang, Emparan and \underline{Virmani} considered several types of potential instabilities, in particular radial perturbations and Gregory-Laflamme \cite{GL} instabilities. By using a heuristic approach based on probing the black ring by removing the balancing condition ---thereby introducing a conical singularity which can be regarded as a force breaking the equilibrium---, it was proven that fat black rings are unstable while thin black rings are stable against radial perturbations \cite{EEV}. On the other hand, from general arguments comparing the relative sizes of the $S^2$ and the $S^1$ sections of the black ring, the instability of thin black rings against a typical Gregory-Laflamme perturbation should be expected. This conclusion was reached in the talk, showing that this kind of perturbations would mostly lead to an eventual fragmentation of thin black rings into a number of black holes. There still exists, however, a small range of values for the angular momentum permitting that the thin black ring be stable against such perturbations \cite{EEV}, though this was unclear at the time.

The NUT 5-dimensional solution was the starting point used by Chen and \underline{Teo} to show how the recently found black holes with ``squashed'' horizons \cite{IM,Wang} can be viewed as a Schwarzschild or Kerr black hole sitting at a fixed point of the NUT spacetime. This fact together with some cleverly devised coordinate transformations allowed the authors to put these solutions in correspondence with the 4-dimensional dyonic static black hole in Kaluza-Klein theory. Turning this correspondence around it was conjectured that the general 5-dimensional Kerr black hole (with two non-vanishing rotation parameters) sitting at a fixed point of the NUT spacetime corresponds to the general rotating dyonic Kaluza-Klein black hole.

\underline{Krtous} {\em et al} studied the separability and integrability, via constants of motion, of the geodesic equations for the general rotating black hole in arbitrary dimensions: the Kerr-NUT-(A)dS spacetime \cite{CLP}. By using the existence of the so-called ``universal'' Yano-Killing tensor \cite{KF} ---universal in the sense that it is independent of the free parameters of the spacetime in suitable coordinates--- the separability of the Hamilton-Jacobi and Klein-Gordon equations was analyzed, and many constants of geodesic motion found. These are quadratic \cite{KKPF} and also of higher order \cite{PKVK}. The geodesic equations were thus shown to be completely integrable \cite{KKPV}. The presentation dealt with the case of an even number of dimensions exclusively, but the results are fully general \cite{PKVK,FKK,KKPF,KKPV}.

\section{Miscellanea}
The last session was not as homogeneous as the previous ones, as it was planned to accommodate the interesting works submitted to the workshop that did not fit in, or have available slots, in the previous ones. 

\underline{Kovar} gave a summary of a number of communications that had been sent to the workshop by the group at the Silesian University in Opava (Czech Republic). This was nicely complemented with a number of interesting posters by members of this group. They have thoroughly considered the dynamics of several test objects near rotating and possibly charged black holes when a positive cosmological constant is assumed \cite{KS,SK,SS}. Equilibrium positions of spinning particles along the axis and on the equatorial plane were found. Similarly, equilibrium configurations of test perfect fluids with a prescribed angular momentum were studied, as well as the velocity profiles of particles and fluids in ``locally non-rotating frames''. Keplerian disks and stable tori were identified, and their dependence on the cosmological constant explicitly shown. The results did not require the existence of an event horizon and the cases with naked singularities were also treated, comparing the corresponding results.

\underline{Dunajski} and Hartnoll presented a talk on gravitational instantons ---solutions of the field equations with Euclidean signature and finite action--- for the Einstein-Maxwell case. The existence of regular multi-centered solutions with arbitrary topology is feasible in this situation ---unlike in the pure vacuum case. The solutions presented in the talk are characterized by having a complex Killing spinor. Furthermore, they were shown to be liftable to (Lorentzian) soliton-type solutions of a 5-dimensional Einstein-Maxwell Chern-Simons theory. These solitons can be regular, without closed causal curves and horizons if they correspond to the 4-dimensional Majumdar-Papapetrou Euclidean solutions \cite{DH}, and then they can be understood as ``solitonic strings''. The possible applications of this work are enumerated in \cite{DH}.

Kagramanova, Kunz and \underline{L\"ammerzahl} presented a description of the timelike and null geodesics in the general static and spherically symmetric line-element written in canonical coordinates. The typical effective potential was identified in general and then this general setting was applied to the solutions of the $SU(2)$ Einstein-Yang-Mills-Higgs theory describing singly charged magnetic monopoles, which are globally regular solutions. The effective potential depends, in this case, not only on the angular momentum but also on the energy of the test particle. The very many qualitatively different orbits were exhaustively studied; for the results and the plots the readers may consult \cite{KKL}.

Elaborating on their previous work concerning classical versus quantum singularities \cite{KH0,KHW}, \underline{Konkowski} and Helliwell studied these matters in cylindrically symmetric spacetimes with some topological defects called dislocations and disclinations. The main conclusion is that all such spacetimes that are classically singular (i.e. causally geodesically incomplete) are also quantum-mechanically singular. The converse also holds  {\em generically} \cite{KH}.

Minguzzi analyzed a very old idea due to Eisenhart \cite{E} relating the spacelike geodesics of a certain $(d+2)$-dimensional spacetime with the trajectories of a classical Lagrangian system with $d$ degrees of freedom. It turns out that the Lorentzian manifold can be characterized by having a covariantly constant null vector field and therefore it belongs to the class found by Brinkmann \cite{B} in 1925 ---though this was not mentioned in the talk. Of course, this class includes $pp$-waves and plane waves \cite{Exact}. The author argued that the geodesics can be chosen to be null or causal and still keep an analogous correspondence, and in fact there is a bijection between null geodesics and the extremals of the Lagrangian. The causal properties of the Brinkmann spacetime were also related with some characteristics of the Lagrangian problem. The final conclusion is that one may try to use techniques of global Lorentzian geometry to tackle unsolved problems in classical mechanics, and the other way round \cite{M}.

The last talk in the workshop was by Knutsen, who criticized the naive but sometimes popular ``derivation'' of the Schwarzschild radius by using the Newtonian spherical potential and the concept of escape velocity. Of course, it is obvious that such a ``derivation'' cannot be minimally rigorous because, to start with, there is no univocal  relation between the curvature coordinate $r$ in the Schwarzschild line-element and the radius in Newtonian physics. Nevertheless, the author went on to prove another inconsistency by extending the ``derivation'' to the Schwarzschild-Kottler solution and comparing the results with the true horizons. Some possible refinements of the ideas behind the ``derivation'' were also put forward. This author also presented a poster where the impossibility of matching a Friedmann-Robertson-Walker dust model with a Schwarzschild exterior at the horizon was deduced. 

\section{Conclusion}
In my opinion, the workshop A1 in Sydney was interesting and informative. I believe that the overall quality of the presentations was very good, and the physics and mathematics of high standards. I am very grateful to all the speakers, and also to the colleagues who attended all or part of the workshop.

\end{document}